\documentclass{osa-article}

\journal{osac}
\usepackage{placeins}
\usepackage{booktabs}
\usepackage{tikz}

\usepackage{enumitem}  % Include the enumitem package

\usepackage{bm}
\usepackage{comment}
\usepackage{subcaption}
\usepackage{cleveref}
\usepackage{amsmath}

\usepackage{multirow}
\usepackage[normalem]{ulem}
\DeclareUnicodeCharacter{2009}{\,}
\usepackage{amsmath}
\usepackage{amsthm}
\usepackage{amssymb}
\usepackage{mathrsfs}
\usepackage{graphics}
\usepackage{hyperref}
\usepackage{cite}
\usepackage{indentfirst}
\usepackage{braket}

\theoremstyle{definition}
\theoremstyle{plain}

\DeclareMathOperator{\Tr}{Tr} %\DeclareMathOperator{\Ext}{Ext}

\newcommand*{\Ketvac}{\Ket{\rm vac}}

\begin{document}

\title{A practical transmitter device for passive state BB84 quantum key distribution}

\author{Yury Kurochkin\authormark{1,*}, Marios Papadovasilakis\authormark{1}, 
Anton Trushechkin\authormark{2}, 
Rodrigo Piera\authormark{1}, 
and James A. Grieve\authormark{1}}

\address{\authormark{1}Quantum Research Center, Technology Innovation Institute, Abu Dhabi 9639, United Arab Emirates\\
\authormark{2}Institute for Theoretical Physics III, Faculty of Mathematics and Natural Sciences, Heinrich Heine University D\"usseldorf, D\"usseldorf 40225, Germany\\
\authormark{*}yury.kurochkin@tii.ae} %% email address is required



\begin{abstract}
In prepare-and-measure quantum key distribution systems, careful preparation of quantum states within the transmitter device is a significant driver of both complexity and cost. Moreover, the security guarantees of such systems rest on the correct operation of high speed quantum random number generators (QRNGs) and the high-fidelity modulation of weak optical signals by high-speed optoelectronic devices, all of which must be hardened against a variety of known side-channel attacks.

A fully passive state preparation approach elegantly resolves these problems by combining state preparation and QRNG stages into a single optical instrument. By using pairs of optical pulses from a gain-switched laser diode as ready-to-use qubits, the QKD transmitter can be radically simplified, eventually comprising a single laser and local phase tomography stage. We demonstrate our simplified transmitter by establishing a QKD link over a 10\,km fiber, generating a secret key rate 110 bits/s, sufficient for practical deployment in ``last mile'' urban quantum networks. Our results show promise in making QKD simpler and more accessible, closing a critical technology gap in building a secure quantum communication infrastructure.
\end{abstract}

\section{Introduction}

\noindent Quantum key distribution (QKD) is a family of communications protocols which exploit the fundamental principles of quantum mechanics to achieve secure key exchange. QKD has attracted significant interest due to its promise of quantifiable security, with proofs based on the laws of physics. Since the first demonstrations, both experimental and theoretical efforts have tended to focus on two goals: extending loss tolerance (and hence distance or range), and increasing the achievable secret key rate. These efforts have pushed the boundaries of quantum technologies and photonics. Today, laboratory demonstrations in fiber can exceed 1000\,km~\cite{TFQKDPan} of range, while commercially available devices are able to produce over 300\,kbps secret key at 100\,km  \cite{toshibaqkd}.

As performance improves, the demands placed upon QKD hardware are also increasing, and a closer look inside QKD devices reveals a wealth of complexity. One key challenge is the preparation of the appropriate quantum states inside the QKD transmitter. Each qubit transmitted in an optimized decoy-state BB84 protocol requires three to five bits of classical randomness \cite{PhysRevA_decoy_Lo}. These bits must be obtained by the use of a quantum random number generator (QRNG), which usually involves the internal preparation and measurement of quantum states, followed by appropriate post-processing. Since today's commercial devices can operate at a rate of more than a Gigahertz, performance requirements for the QRNG component are substantial. Once these classical bits are obtained, they are used to set the state of electro-optical modulators, which ultimately define the quantum state of the transmitted qubit. The modulators are necessarily high-end devices, as even small deviations from the intended modulation can lead to potential security vulnerabilities. For example, a small voltage correlation between consecutive signals might be acceptable in conventional communications as long as the 0 and 1 state can be distinguished, but such a correlation would be catastrophic for a QKD device. Indeed there is a small library of known side-channel attacks related to the active modulation system (including e.g. so-called Trojan horse attacks~\cite{gisin2006trojan}), and it is common for QKD devices to contain multiple counter-measures designed to harden the system \cite{Makarov_countermeasures}. As well as complicating practical assessments of security, this ``active-modulation'' approach drives system cost, with the result that today's QKD hardware remains comparatively expensive.

In this work, we take a different path. Instead of relying on a complicated two-stage preparation procedure, we advocate a direct approach: the passive state preparation method. More than 30 years ago, Artur Ekert proposed to use entanglement to generate quantum keys \cite{Ekert_1991}. An enduring advantage of this method is the absence of a QRNG, with entropy entering the system due to the inherent randomness of projected measurements on entangled states. Passive state preparation builds on this idea, using natural quantum randomness such as the random phase of the gain switch laser to perform state preparation directly. This randomness is widely used today in both laboratory and commercial QRNG devices~\cite{Abellan_QRNG, Kurochkin_QRNG}. In 2010, Curty et~al. proposed its use in the preparation of polarization encoded qubits~\cite{PassiveBB84_Curty_2010}, and it has more recently attracted attention for its safety advantages~\cite{PassiveQKD_Wang_2023}.

At the core of this proposal is the combination of two coherent states with random phase on a polarizing beam-splitter, obtaining a random polarization at the output. One can then split the signal into a weak coherent pulse (WCP) sent to the receiver and a bright classical signal sent to a local tomography device (within the transmitter). This tomography allows to exclude all states except the four BB84  states~\cite{BB84}, either by use of a fast switch or simply by post-selection. Later, this approach was used for a fully passive decoy state BB84~\cite{FullyPassive4_Wang} requiring four lasers ideally tuned to indistinguishability. To overcome wavelength matching problems, experimental demonstrations later replaced four lasers by one laser and a fast optical switch \cite{ExpPassiveQKD_Lu_2023}. A passive state generation approach with laser interference was also experimentally shown to generate the random amplitude required for decoy states \cite{liu2017practical}.

We set out to build on this work, with an attempt to construct the ``simplest possible'' QKD transmitter, containing the minimum number of active optical elements. By pursuing this goal we hope to both reduce the cost associated with QKD hardware, and increase the practically achievable security by eliminating many components which commonly contribute to side-channel attacks.

In our system, a pair of consecutive pulses from a gain-switched laser diode are deployed as a ``ready-to-use'' qubit, reducing the number of active elements to a single laser. Local tomography is simplified to a single fiber loop, which converts phase difference into polarization. Finally, a two-basis polarimeter completes the scheme.
While this approach does not naturally lend itself to high key rates or long distances, we believe that it may hold the key to making QKD more affordable for wider use. We further simplify by discarding the decoy state protocol and mitigating the photon number splitting attack via privacy amplification, resulting in loss tolerance of up to 10\,dB. With this device, we demonstrate a quantum bit error ratio (QBER) of 5.4\% at a distance of 10\,km, and an secret key rate of 110 bps (under asymptotic conditions). We believe this performance is sufficient to realize simple and low-cost QKD devices for ``last-mile'' urban networks, a missing technology building block for the quantum networks in development today.

\section{Experimental setup}

\begin{figure}[!ht]
\centering\includegraphics[width=0.9\linewidth]{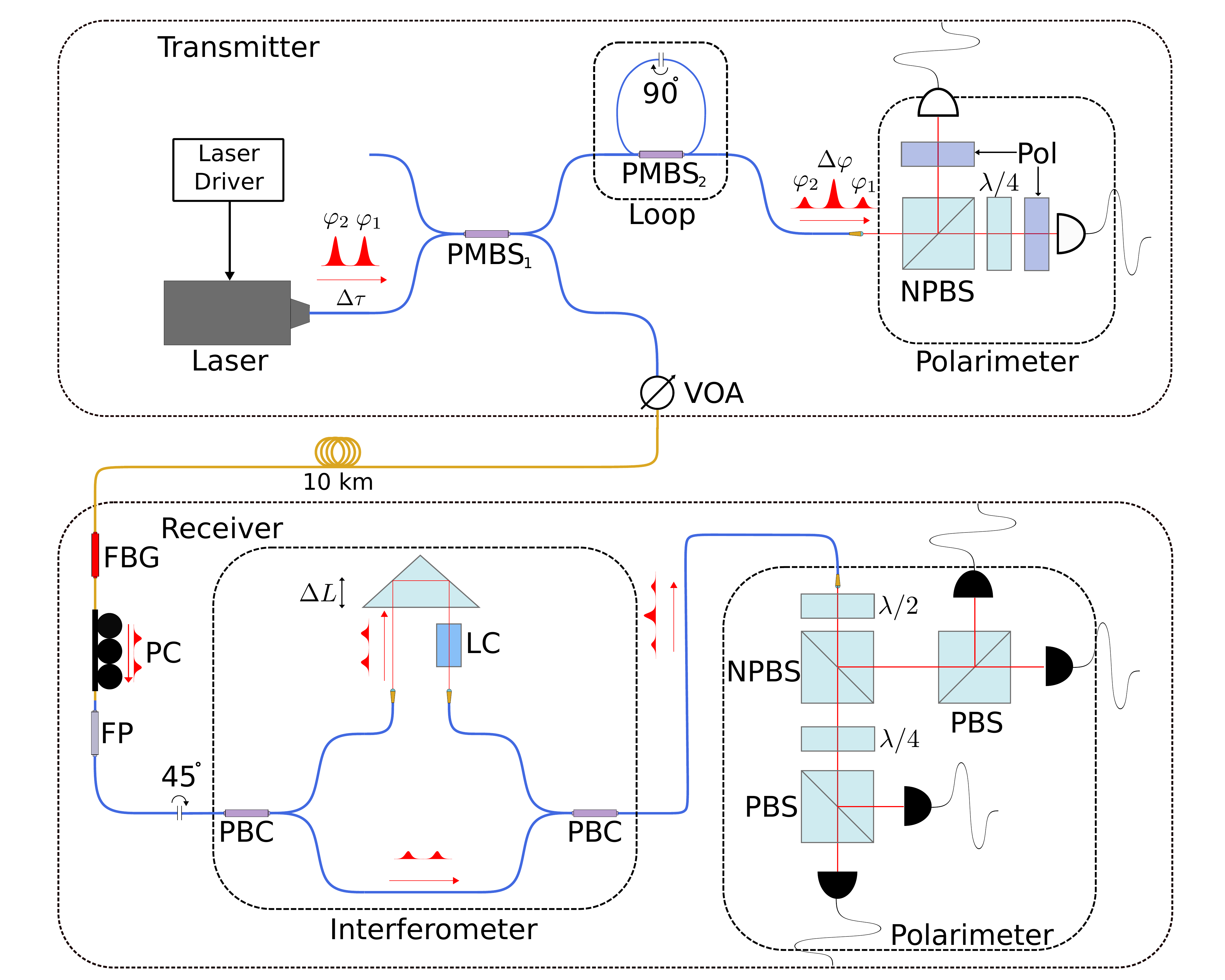}    
\caption{
Experimental setup: A DFB laser produces a pulse pair with random phases $\varphi_1$ and $\varphi_2$ sent through a PMF (Blue Line). A Polarization-maintaining fiber Beamsplitter ($PMBS_{1}$) splits the signal into two outputs, one sent to the Polarimeter through the tomography channel (top output) and the other sent through the quantum channel (bottom output). In the Tomography channel, the random phase difference $\Delta\varphi=\varphi_1 - \varphi_2$  is mapped to a polarization state utilizing a loop consisting of $PMBS_{2}$ and $90^{\circ}$ rotator connecting the two principal axes of fiber. A free space polarimeter splits signal using a nonpolarizing beamsplitter (NPBS) and measures circular (diagonal) output polarization with (without) a quarter wave plate $\lambda/4$ followed by polarizer (Pol) and amplified detector (represented by a white semicircle). In the Quantum Channel, the light is attenuated by a VOA to a small fraction of the signal and then sent to the receiver through 10 km lengths SMF (yellow line). At the receiver, the signal is filtered by Bragg filter FBG, and the polarization is compensated using a PC. Fiber polarizer (FP), $45^{\circ}$ rotator connector followed by a PBC allows to split the signal and combine it into an imbalance Mach-Zehnder interferometer. A tunable delay $\Delta L$ allows to match the delay with the delay into the transmitter loop. A Liquid crystal (LC) compensates for the relative temperature drift between the Receiver interferometer and the Transmitter loop. At the output of the PBC, the phase difference is mapped to polarization  and is measured in the projection system, which consists of a free space NPBS, PBS, waveplates ($\lambda/2$, $\lambda/4$), and SNSPD (represented by black semicircles)}
\label{fig:experimental_setup.png}
\end{figure}

\noindent To generate time bin qubits with random phase, the transmitter (Alice) uses a 1310\,nm DFB laser diode operated in ``gain switched'' mode. To obtain the required phase randomness via the ``phase diffusion'' process \cite{abellan2014ultra, PhysRevA_107_012616}, we set the voltage between pairs of electrical drive pulses to zero and check the interference pattern. Using a bench-top pulse generator, we can achieve 500\,ps optical pulses with $\Delta t=18.9\,ns$ delay and $f=1.5\,MHz$ repetition rate. The experimental setup is shown in Fig. \ref{fig:experimental_setup.png}. The optical signal is split by an on-fiber Beam Splitter (FBS) into the quantum (bottom output) and tomography (top output) channels. The tomography signal, sent through the tomography channel, is measured locally. A common technique for measuring time-bin qubits is the unbalanced Mach-Zehnder interferometer \cite{PhysRevA_89_062328}. An interferometer can distinguish between two orthogonal states, for example $\Delta \varphi = 0$ and $\Delta \varphi = \pi$. However, the same interferometer would not distinguish the second orthogonal basis: $\Delta \varphi = \pi/2$ and $\Delta \varphi = 3\pi/2$. One can use two interferometers, but then it would be challenging to stabilize the relative phase between the two postselected bases, and this would increase the complexity of the transmitter.

A practical solution is to map the phase difference into the polarization domain. To minimize the number of elements in the transmitter, we introduce a delay line with only one beamsplitter and one loop of polarization-maintaining fiber (PMF). Inside the loop, a $90\deg$ rotating connection couples light from the slow axis to the fast axis. For simplicity, the polarization along the slow axis is referred to as vertical and along the fast axis as horizontal. The loop delay line converts the first pulse (early) of the pair into a horizontal polarization component and the second pulse (late) into a vertical polarization component of the output pulse. The polarization state of this combined signal depends on the phase difference between the horizontal and vertical components. 
 
\begin{figure}[t]
\centering
\begin{minipage}{0.35\linewidth}
    \includegraphics[width=\linewidth]{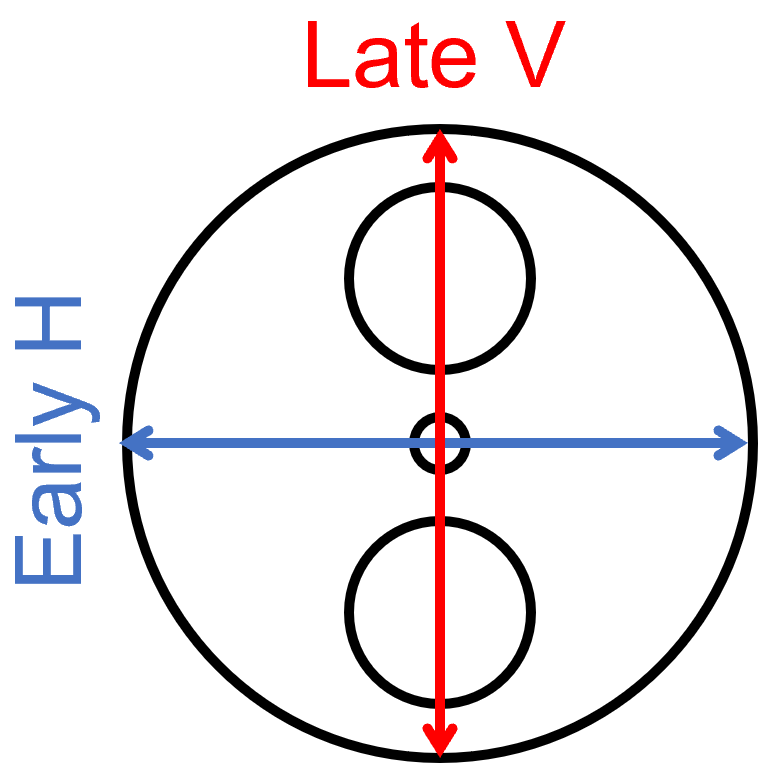}
    %\subcaption{}
    %\label{fig:pulses_polarization_a}
\end{minipage}\hfill
\begin{minipage}{0.5\linewidth}
    \includegraphics[width=\linewidth]{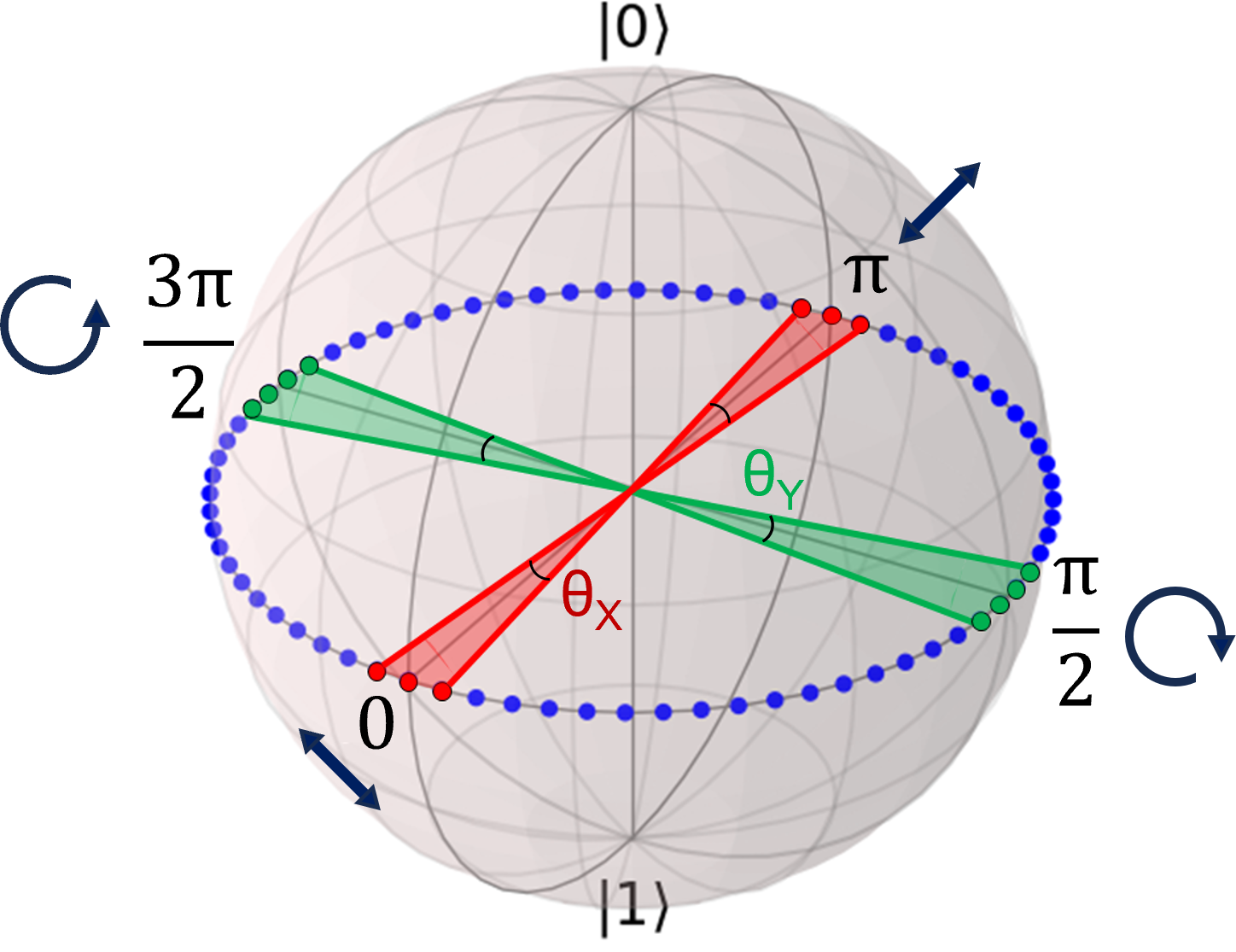}
    %\subcaption{}
    %\label{fig:pulses_polarization_b}
\end{minipage}
\caption{(Left) Output of the polarization maintaining fiber where different time pulses are converted to different polarization components aligned with fiber principal axes. (Right) Correspondence between: laser output pulses phase difference ($\Delta\phi=$$0, \pi/2, \pi, 3\pi/2$), loop output polarizations ($\mathrel{\text{$\nearrow$\llap{$\swarrow$}}}$, $\mathrel{\text{$\nwarrow$\llap{$\searrow$}}}$, $\circlearrowleft$, $\circlearrowright$,) and qubit states as postselected Bloch sphere equator regions.}
\label{fig:pulses_polarization}
\end{figure}

Put explicitly, for a phase difference $\Delta \varphi = 0$, the polarization is diagonal. For $\Delta \varphi = \pi$ the polarization is antidiagonal and for $\Delta \varphi = \pi/2$ and $\Delta \varphi = 3\pi/2$ the polarization is circular/countercircular (see Fig.~\ref{fig:pulses_polarization}). To distinguish between the polarizations, the light is directed into a free-space polarimeter comprising a linear polarizers and photodiodes. A single polarizer-photodiode pair can distinguish between two linear polarizations by measuring a voltage amplitude. To distinguish circular/countercircular polarizations we use a $\lambda/4$ waveplate to rotate the light entereing one such pair. In this case, the precision of the $\lambda/4$ waveplate determines the phase difference between two bases. The loop delay line also creates unmatched early and late signals that must be discarded, but these do not create a security concern as they are well isolated from the external channel.

The polarimeter outputs, i.e. the analog voltage signal of the photodiodes, are measured by a time tagger (TT, Swabian Instruments). The time tagger input comparator is tuned to detect only the maximum or minimum of the tomography output. These comparator levels define a region on the Bloch sphere equator, which represents the postselection criteria (see Fig.~\ref{fig:pulses_polarization} (Right)).

The quantum signal, sent through the quantum channel, is attenuated to a weak coherent state by a variable optical attenuator (VOA). The amplitude of this state is set to approximately $\mu = 0.15$ photon/pulse, which is enough to take into account the photon number splitting attack under trusted receiver loss. The signal is then sent to the receiver (Bob) via a 10\,km long Single-Mode Fiber (SMF) (G.652 specification) deployed in a metropolitan fiber conduit.

At the receiver, the signal is spectrally filtered by a fiber Bragg grating (FBG), followed by a polarization controller and fiber polarizer (FP) to linearize the polarization of the input signal. A $45^{\circ}$ rotating connector followed by a polarization beam combiner (PBC) splits the signal onto an imbalanced Mach-Zehnder interferometer (MZI) input. The long arm of the MZI imposes a $\sim\Delta L$ delay, (time delay of $\sim\Delta t$), and an active free-space phase modulator comprising a liquid crystal (LC). This modulator allows us to tune the relative phase between the delay lines of Alice and Bob, with the receiver effectively actively tracking the (slowly drifting) phase of the transmitter.

The unbalanced MZI output maps the two temporal modes to two polarization modes, which are then measured in the projection system. The subsequent BB84 state projective system has four outputs: $X_1$ (diagonal-D), $X_0$ (antidiagonal-AD), $Y_1$ (right circular-R) and $Y_0$ (left circular-L). 
When the relative phase is correctly tuned, the two channels of the transmitter tomography correspond to the two bases at the receiver. Coincident signals between the postselection events of the transmitter and receiver single photon detectors enable the distillation of a sifted key.

\section{Key generation Theory}

\noindent In this section, we describe the security model used in this quantum key distribution scheme, with a focus on those adaptations made to accommodate probabilistic state preparation with finite precision.

Beginning with the transmitter, we denote $I$ the intensity of the pulses entering the tomography module and consider a pulse with the intensity $\frac{I}{2}|1+e^{i\Delta\varphi}|^2=I(1+\cos\Delta\varphi)$ entering detector~X. By establishing upper and lower thresholds $I^{\rm high}$ and $I_{\rm low}$, we can postselect pulses with phase difference $|\Delta\varphi|\leq\delta\varphi$ and $|\Delta\varphi-\pi|\leq\delta\varphi$. Analogously, on detector~Y, we postselect pulses with phase difference $|\Delta\varphi\pm\frac{\pi}2|\leq\delta\varphi$. Note that we neglect intensity fluctuations in the laser (which we leave for future works).

The analysis generally follows that from Ref.~\cite{PassiveDecoy_Zapatero_2023}. Denote the $\mu/2$ the intensity of each of two pulses entering into the quantum channel. The phase-randomized quantum state before the postselection is
\begin{equation}
\label{eq:rho}
    \rho=\frac1{4\pi^2}
    \int_0^{2\pi}\!d\varphi_1\!
    \int_0^{2\pi}\!d\varphi_2
    \,
    P\bigg[\bigg|\sqrt{\frac{\mu}2} e^{i\varphi_1}\bigg\rangle_1
    \bigg|\sqrt{\frac{\mu}2} e^{i\varphi_2}\bigg\rangle_2\bigg]
    =
    \frac1{2\pi}
     \sum_{n=0}^\infty
    \frac{\mu^n}{n!}e^{-\mu}
   \int_0^{2\pi}d\Delta\varphi\,
    P[{\ket n}_{\Delta\varphi}],
\end{equation}
where $P[\ket\psi]$ denotes the projector onto the vector $\ket\psi$, ${\ket\alpha}_{1,2}$ for a complex number $\alpha$ denotes the corresponding coherent state for two time bins. Further, 
\begin{equation}
    {\ket n}_{\Delta\varphi}=\frac{(a_{\Delta\varphi}^\dag)^n}{n!}\Ketvac,
    \qquad
    a_{\Delta\varphi}^\dag=\frac{a_1^\dag+e^{i\Delta\varphi}a_2^\dag}{\sqrt2},
\end{equation}
where $\Ketvac$ is the vacuum vector, and $a_{1,2}^\dag$ are creation operators for two time bins, respectively.

After the postselection of the phase near the value $x\in\{0,\pi,\frac\pi2,-\frac\pi2\}$, the unnormalized state is
\begin{equation}
    \rho_x=\frac1{2\pi}
     \sum_{n=0}^\infty
    \frac{\mu^n}{n!}e^{-\mu}
    \int_{x-\delta\varphi}^{x+\delta\varphi}P[\ket{n}_{\Delta\varphi}]\,d\Delta\varphi
\end{equation}
and the probabilities of the postselection of each $x$ are the same and equal to $p_x=\Tr\rho_x=\delta\varphi/\pi$. We can use the standard GLLP \cite{GLLP} formula for the secret key rate
\begin{equation}
    r=Y_1[1-h(e^{\rm ph}_1)]-\text{leak}_{\rm EC},
\end{equation}
where $Y_1$ is the ratio of positions obtained from the single-photon ($n=1$) state, $e^{\rm ph}$ is the fraction of phase errors in the single-photon states, $h$ is the binary entropy function, and $\text{leak}_{\rm EC}$ is the number of bits of information announced during the error correction procedure.

We have
\begin{equation}
    Y_1=1-\frac{p_{\rm multi}}Q,
\end{equation}
where
$p_{\rm multi}=1-(1+\mu)e^{-\mu}$ is the probability of a multiphoton pulse and $Q$ is the total gain. 

Now we need to estimate $e_1^{\rm ph}$. The normalized single-photon contribution to $\rho_x$ can be expressed as

\begin{equation}
    \rho_x=\frac1{\delta\varphi}
    \int_{x-\delta\varphi}^{x+\delta\varphi}\ket{1}_{\Delta\varphi}\bra{1}\,d\Delta\varphi 
    =
    \frac{1+\Delta}2
    \ket{x}\bra{x}
    +
    \frac{1-\Delta}2
    \ket{x+\frac\pi2}\bra{x+\frac\pi2},
\end{equation}
where $\Delta=\frac{\sin\delta\varphi}{\delta\varphi}$ and $\ket{\phi}=\frac1{\sqrt2}(a_1^\dag+e^{i\phi}a_2^\dag)\Ketvac$ for an arbitrary $\phi$. Then, completely analogously to Ref.~\cite{PassiveDecoy_Zapatero_2023}, one can show that the phase error rate in each basis is (asymptotically) equal to the phase error rate in the conjugate basis. Hence, if we average over two bases, then the bit error rate is (asymptotically) equal to the phase error rate. The single-photon bit error rate can be upper bounded by $\rm{QBER}/Y_1$, where QBER represents the total QBER. Thus, we arrive at the final formula for the secret key rate
\begin{equation}
    r=Y_1
    \Big[
    1-h\Big(
    \frac{\rm QBER}{Y_1}
    \Big)\Big]
    - {\rm leak}_{\rm EC}.
\end{equation}

\section{Results}

\begin{figure}[!ht]
\centering\includegraphics[width=0.5\linewidth]{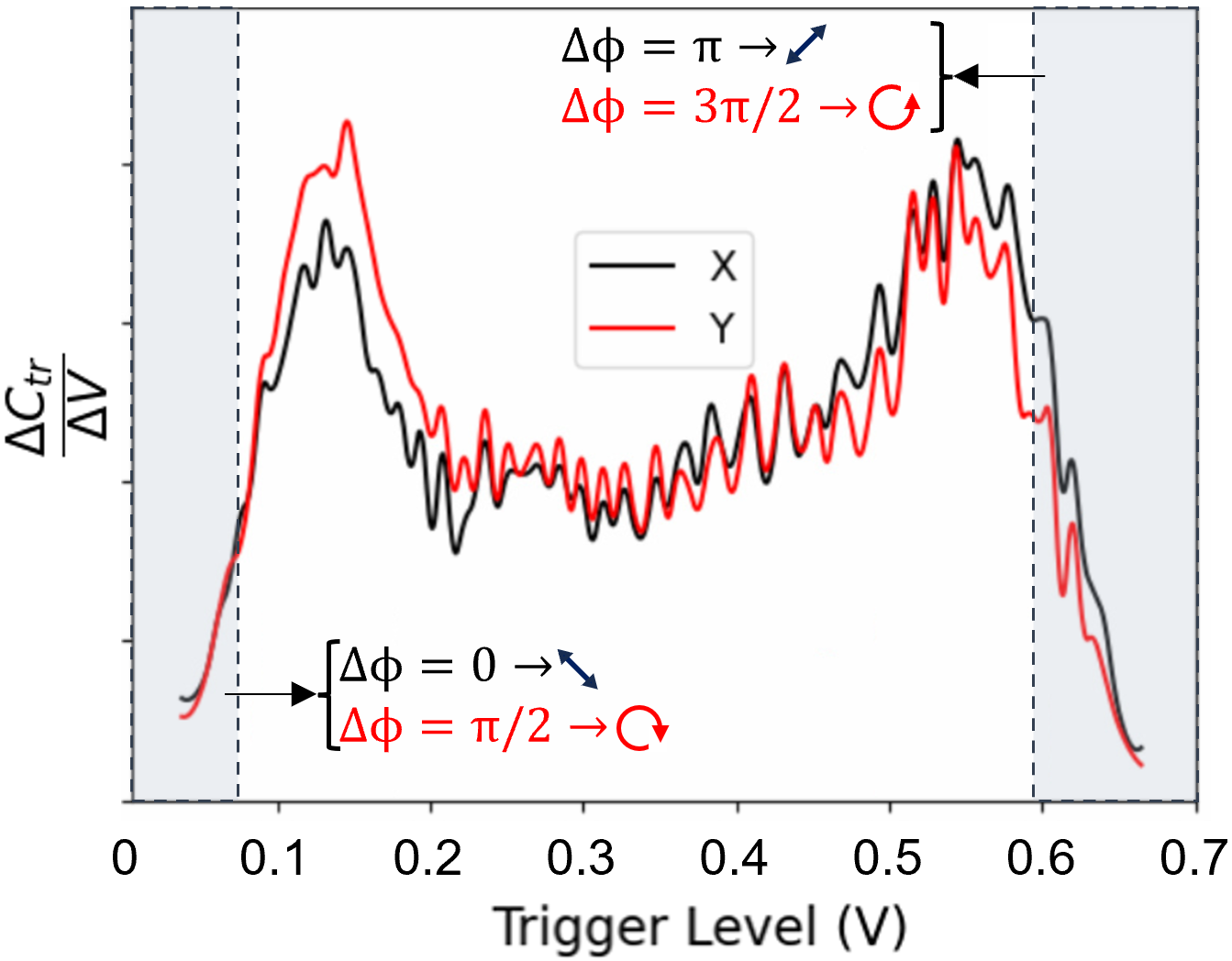}
\caption{Phase diffusion diagram measured by Transmitter. The grayed-out regions define the post-selected state}
\label{fig:phase_diffusion}
\end{figure}

\noindent To make passive state preparation valid one needs to verify phase randomness quality by measuring the (now standard) arcsine phase diffusion histogram \cite{PhaseDiffusion_Mitchell_2015, Chirp_Kurochkin}, which is inherent to a uniformly random distribution of $\Delta\phi_0$ Fig.~\ref{fig:phase_diffusion} shows representative data from our system, with the arcsine relation observed for both bases separated by a $\pi/2$ rotation (X and Y). In addition to confirming the randomness of the initial phase of the generated laser pulses, the figure can be used to visalize the postselection strategy, with the two shaded regions denoting typical upper and lower trigger levels selected for state preparation.

\begin{figure}[ht]
\centering
\begin{minipage}{0.5\linewidth}
    \includegraphics[width=\linewidth]{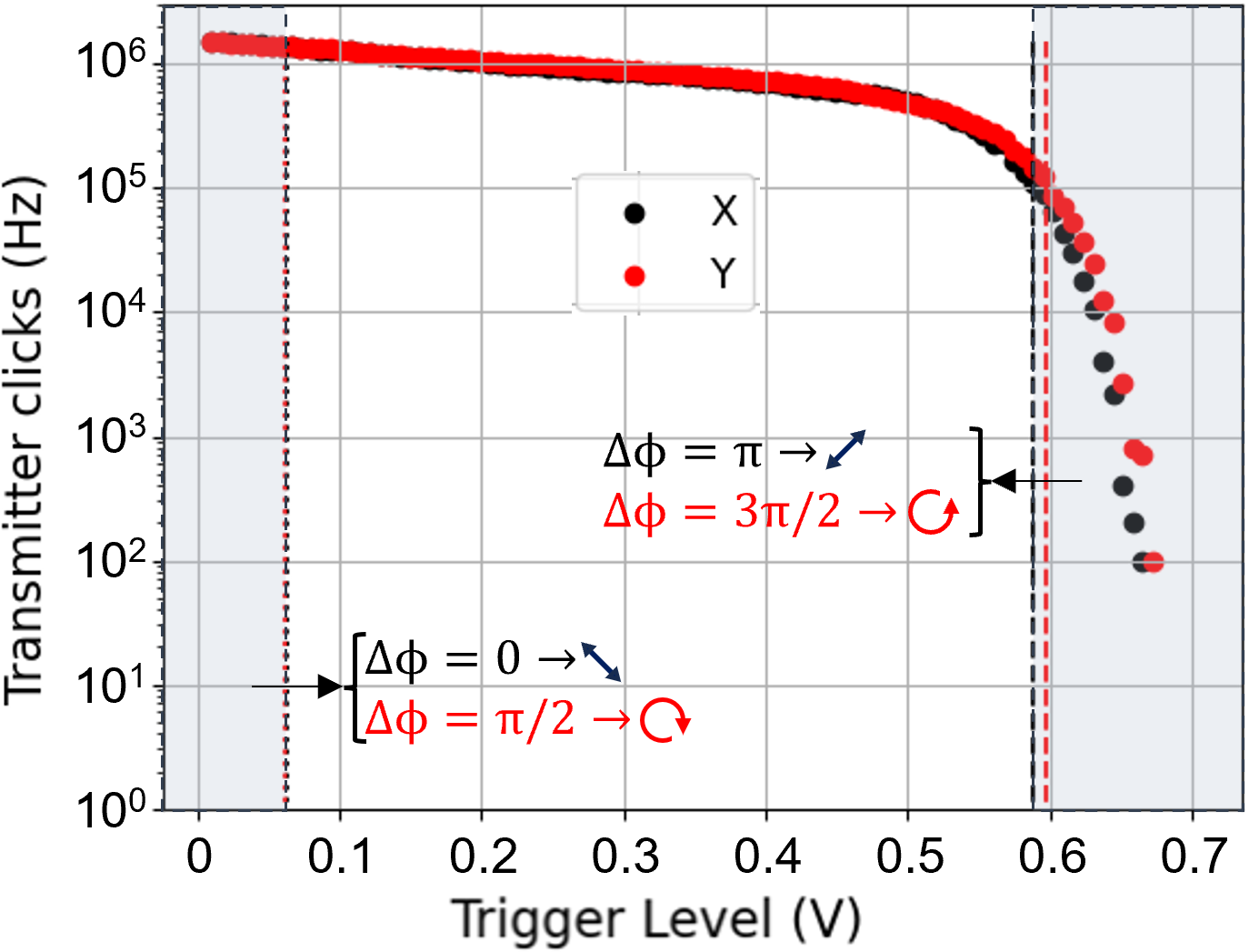}
    %\subcaption{}
    %\label{fig:pulses_polarization_a}
\end{minipage}\hfill
\begin{minipage}{0.5\linewidth}
    \includegraphics[width=\linewidth]{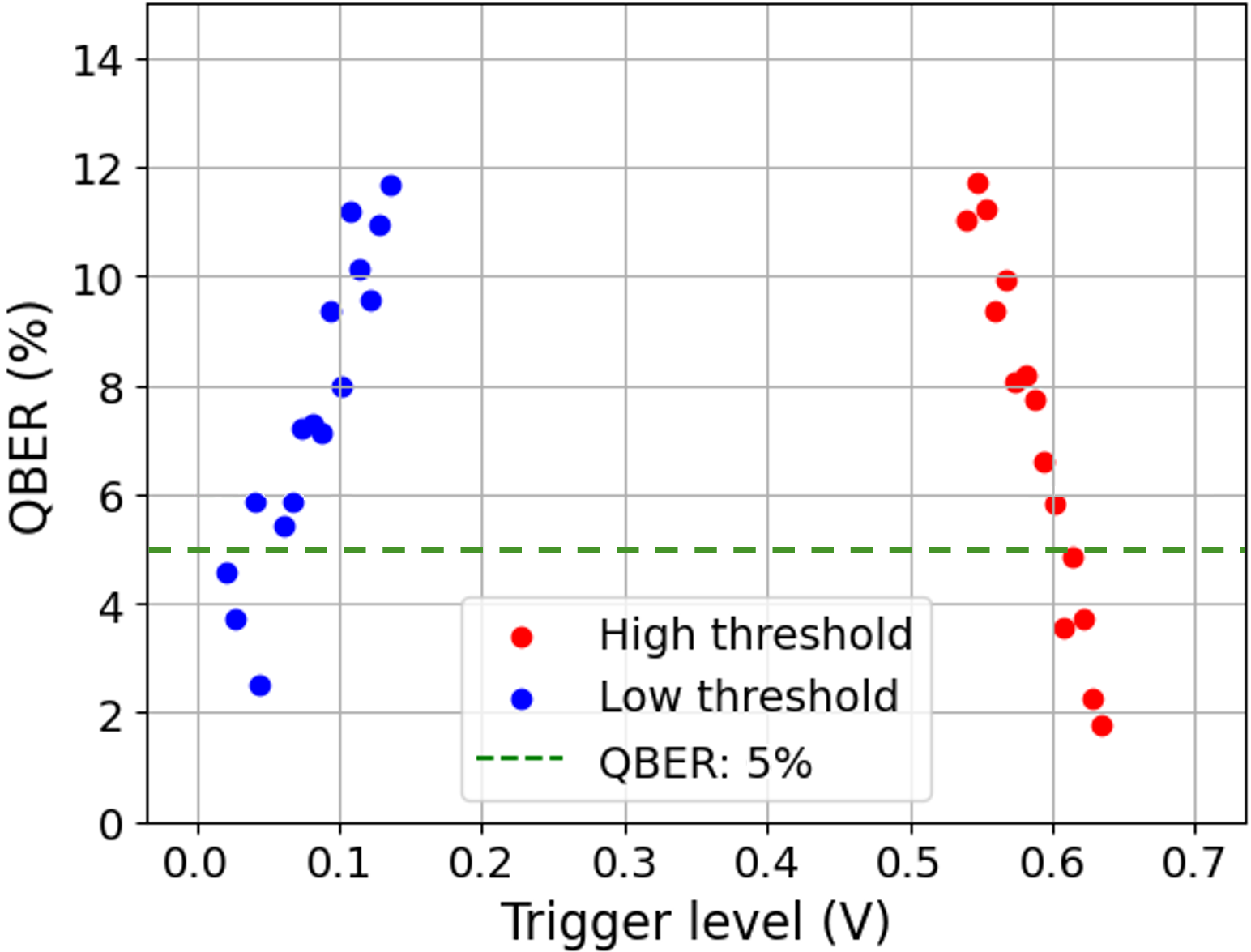}
    %\subcaption{}
    %\label{fig:pulses_polarization_b}
\end{minipage}
\caption{(Left) Transmitter diagnostics of the $X$ basis. (a) transmitter clicks as a function of trigger voltage level. The dashed lines correspond to the trigger levels, which post-select roughly 10\% (75 kHz for each basis) of the total generated states for high, and low thresholds; (Right) Tabletop measured QBER for high and low threshold conditions, as a function of TT trigger level. The QBER is decreased as the post-selection conditions become more stringent}
\label{fig:postselection_rate}
\end{figure}

A critical step in passive state preparation is the the adjustment and optimization of postselection comparator level.
Fig.~\ref{fig:postselection_rate}(Left) shows the relationship between the high-side postselection criterion and the observed success rate at the transmitter (``clicks''), which decrease with increasing comparator level. It is intuitive that tight postselection criteria lead to more precice state preparation and and consequently lower QBER (see Fig.~\ref{fig:postselection_rate}(Right)). In this work we select postselection criteria leading to QBER around $5\%$, though in a deployed QKD system this could be determined by a proper optimization process running against the final key rate.

At the output of the transmitter, the mean photon number is chosen for each distance. For 11.58\,km deployed fiber (6.7\,dB loss) it is set to $0.15$ photon/qubit (where the qubit is defined over a pair of time bins) using a variable optical alternator (VOA). This value is chosen as a trade-off between obtaining sufficient signal at the receiver, while maintaining only moderate key loss due to the assumed presence of the photon number splitting attack.

Our observed sifted key rate is 2.5\,kbit/s with $5.8\%$ QBER, this yields 15\,bit/s secret key.

On the receiver side we tune the imbalanced MZI to minimize QBER, with active phase-tuning achieved using a liquid crystal (LC). Dynamic variations of the phase  are caused by thermal variation along the delay lines in both transmitter and receiver. During phase tuning operations, the transmitter announces all prepared states until the receiver confirms low enough QBER to switch to key generation mode. We re-enter phase tuning mode switches frequently enough to keep average QBER abouit $5\%$. About $50\%$ of the time we are stabilizing MZI phase what leads to additional key rate decrease. FPGA based electronics will allow to reduce MZI stabilization time in the future.

\begin{figure}[t]
\centering\includegraphics[width=1\linewidth]{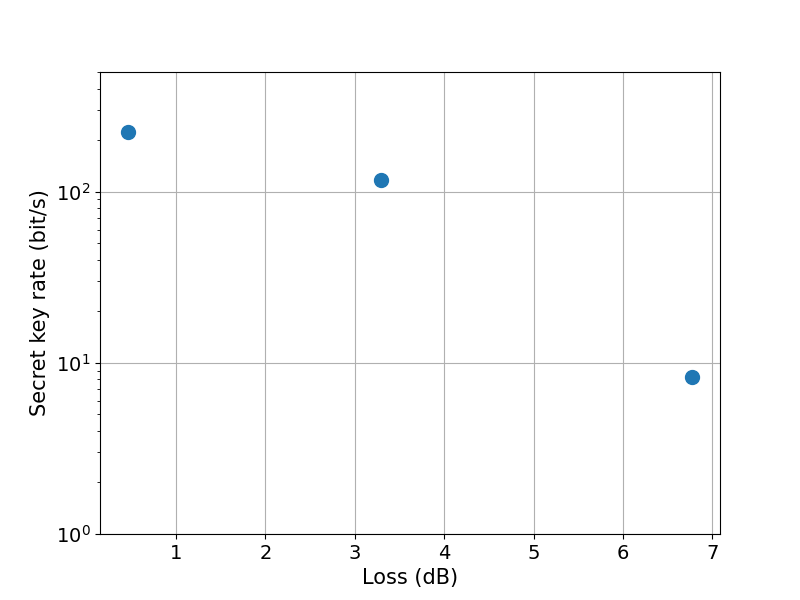}
\caption{Secret key rate for various quantum channel losses. Due to the high detector quality, the dark counts are negligible. Measured QBER is mainly due to the non-ideal state preparation and optical misalignment. Quantum channels used in the experiment: 2 m fiber, 10 km fiber spool and 11.58\,km deployed fiber.}
\label{fig:lc_adjust}
\end{figure}

\section{Conclusions}
%V3
\noindent In our study, we have effectively demonstrated a simplified ``last-mile'' QKD scheme tailored for urban networks. By using a passive state preparation method, our system overcomes the complexity and cost associated with traditional QKD devices while minimizing vulnerability to side channels. With a key rate of 10-100 bits/s over 10 km, this system is compatible with the current IT infrastructure where 256-bit quantum keys are typically uploaded every minute or more. The addition of an optical switch could further improve connectivity and connect multiple users to a single quantum network node. This development is an important step towards the practical deployment of secure quantum communication in data centers and urban networks and could make QKD technology affordable for end users.

%\bibliography{QKD}
% \printbibliography

\end{document}